\RequirePackage{fix-cm}
\documentclass[smallextended]{svjour3}       
\smartqed  

\usepackage{graphicx}
\usepackage[most]{tcolorbox}
\usepackage{booktabs}
\usepackage{multirow}
\usepackage{tabularx}
\usepackage{threeparttable}
\usepackage{appendix}
\usepackage{pifont}
\usepackage{soul}
\usepackage{url}
\usepackage{hyperref}
\newcommand{\sectopic}[1]{\vspace{0.2em}\par\noindent{\textit{\bfseries #1}}}

\begin{document}

\title{A Multi-solution Study on GDPR AI-enabled Completeness Checking of DPAs}


\author{Muhammad Ilyas Azeem\textsuperscript{a}  \and
        Sallam Abualhaija\textsuperscript{a} \and
}


\institute{\at
              \textsuperscript{a}SnT, University of Luxembourg, Luxembourg \\
              \email{\{ilyas.azeem, sallam.abualhaija\}@uni.lu}           
}

\date{Received: date / Accepted: date}

\maketitle

\begin{abstract}
Specifying legal requirements for software systems to ensure their compliance with the applicable regulations is a major concern to requirements engineering (RE). Personal data which is collected by an organization is often shared with other organizations to perform certain processing activities. In such cases, the General Data Protection Regulation (GDPR) requires issuing a data processing agreement (DPA) which regulates the processing and further ensures that personal data remains protected. 
Violating GDPR can lead to huge fines reaching to billions of Euros. 
Software systems involving personal data  processing must adhere to the legal obligations stipulated in GDPR and outlined 
in DPAs. Requirements engineers can elicit from DPAs legal requirements for regulating the data processing activities in software systems. Checking the completeness of a DPA according to the GDPR provisions is therefore an essential prerequisite to ensure that the elicited requirements cover the complete set of obligations. Analyzing DPAs with respect to GDPR entirely manually is time consuming and requires adequate legal expertise. 
In this paper, we propose an automation strategy that addresses the completeness checking of DPAs against GDPR provisions as a text classification problem. Specifically, we pursue ten alternative solutions which are enabled by different technologies, namely traditional machine learning, deep learning, language modeling, and few-shot learning.  The goal of our work is to empirically examine how these different technologies fare in the legal domain. We computed F$_2$ score on a set of 30 real DPAs. Our evaluation shows that best-performing solutions yield F$_2$ score of 86.7\% and 89.7\% are based on pre-trained BERT and RoBERTa language models. Our analysis further shows that other alternative solutions based on deep learning (e.g., BiLSTM) and few-shot learning (e.g., SetFit) can achieve comparable accuracy, yet are more efficient to develop.

\keywords{Requirements Engineering (RE) \and The General Data Protection Regulation (GDPR) \and Regulatory Compliance \and Data Processing Agreements (DPAs) \and Artificial Intelligence (AI) \and Natural Language Processing (NLP) \and Classification \and Large Language Models (LLMs) \and Few-shot Learning (FSL) \and Data Augmentation.}
\end{abstract}

\section{Introduction}
\label{sec:introduction}

\textit{Legal requirements} (also known as \textit{compliance requirements}) describe the behavior and functions of a software system pursuant to applicable regulations~\cite{Maxwell2012}. Developing compliant software systems requires the elicitation of legal requirements from regulations, an essential activity in the requirements engineering (RE) field. Both manual and automated approaches have been investigated for navigating through the regulations to create machine-analyzable representations and extract compliance-relevant information (e.g.,~\cite{Breaux:22,Okhaide22,Torre:20Sosym,Soltana:18,Ghanavati2014}). Despite the community efforts, software development practices are still introducing personal data breaches, e.g., unauthorized sharing with third-parties~\cite{Feal2021}. One of the main challenges for properly implementing legal requirements in software systems is due to the complexity of the legal language used in regulations. 
Eliciting legal requirements is time-consuming and error-prone and often requires adequate legal expertise. 
For example, the General Data Protection Regulation (GDPR), enforced in the European Union (EU) in 2018~\cite{GDPR2018}, contains provisions on data privacy and security with which organizations, located inside or outside Europe, must comply as long as they collect and process personal data of people in the EU. Until today, organizations are clearly struggling in understanding how to comply with GDPR considering the fines being imposed yearly due to different types of breaches, most of which are caused by non-compliant practices in the software systems deployed by organizations. Statistics show that most violations are related to breaching data processing principles, leading to 418 fines of more than 1,500 billions of euros\footnote{\href{https://www.enforcementtracker.com/?insights}{https://www.enforcementtracker.com/?insights}}.     
Eliciting legal requirements for a specific application context by navigating through the 88 pages of GDPR with its 173 recitals and 99 articles divided into 11 chapters is a daunting task for requirements engineers.

To bridge the legal knowledge gap and optimize the elicitation of legal requirements, we argue in this paper that a \textit{regulated document} (e.g., privacy policy) provides compliance-relevant information which can be easier to navigate than the entire regulation. Our work focuses on data processing agreements (DPAs), which are legal contracts between a \textit{controller} who collects personal data and a \textit{processor} who processes personal data on the controller's behalf~\cite{pantlin:18,amaral:21}. 
To be an effective source document for eliciting legal requirements, DPAs must be GDPR-compliant, meaning that the content of a DPA must be \textit{complete} according to the provisions of GDPR. Otherwise, the resulting legal requirements elicited from the DPA will be incomplete. 
Verifying the completeness of DPAs against GDPR is thus a pre-requisite for leveraging the DPAs as source knowledge for eliciting legal requirements for software systems that process personal data. 

\begin{figure} [t!]
\centering
\includegraphics[width=.98\textwidth]{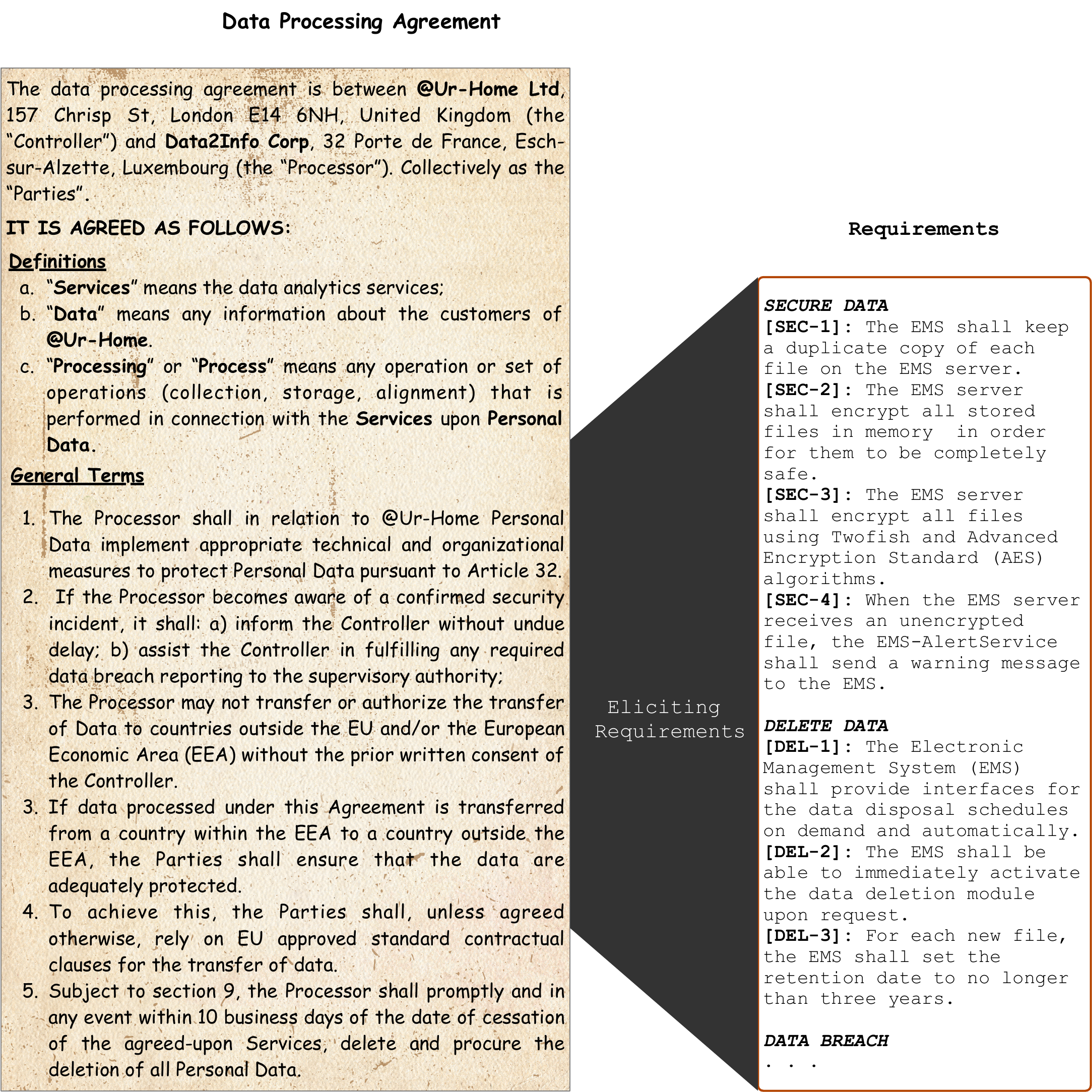}
\caption{Example of eliciting requirements from a DPA.}
\label{fig:example}
\end{figure}

Fig.~\ref{fig:example} shows an example of eliciting technical requirements from the   legal statements in a given DPA. On the left side of the figure, we show an excerpt from a GDPR-compliant DPA, and on the right side we show some requirements that can be elicited from the DPA. 
The DPA is between \textit{@Ur-Home Ltd}, a large e-commerce company selling various products, and \textit{Data2Info Corp}, a company that provides data analytics services. The former acts as the controller and the latter as the processor. 
The \textit{Data2Info} company uses an electronic management system (EMS) for managing files. To comply with GDPR, the EMS must, among other things, incorporate security mechanisms to protect personal data and further delete all personal data received form the controller upon the termination date of the agreement. The first four requirements correspond to additional security mechanisms that need to be accounted for when storing files on the server. For instance, if a file is not encrypted, the user shall be immediately informed. The last three requirements are about the deletion of the data. Relying on an incomplete DPA for eliciting requirements entails the risk of missing technical requirements which might be critical for compliance with GDPR. Such cases can be found when the system makes a failure, e.g., personal data transferred without sufficient protection mechanisms are prone to breach incidents. The defect of missing requirements can only then be detected when it is too late. Since the system failure in relation with compliance can lead to large fine\footnote{\href{https://www.cnbc.com/2023/05/24/irish-data-regulator-defends-1point3-billion-meta-fine.html}{https://www.cnbc.com/2023/05/24/irish-data-regulator-defends-1point3-billion-meta-fine.html}}.

Checking the completeness of DPAs against GDPR has been investigated in a recent work by Amaral \textit{et al.}~\cite{amaral:22}. Amaral \textit{et al.} define, in collaboration with legal experts, a set of 45 GDPR provisions that regulate the content of a DPA. To conclude whether a DPA is complete according to GDPR, the authors develop a rule-based automated approach that verifies the textual content of a given DPA against the GDPR provisions. Their approach utilizes natural language processing (NLP) technologies to analyze the semantic similarity of the textual content in a DPA against GDPR. While their approach has demonstrated its effectiveness on checking the completeness of real DPAs, using rules has several drawbacks. First, regulations are subject to continuous changes which can impact, to a large extent, the compliance checking process~\cite{johansson2019,breitbarth2019}. 
Adjusting rules to cover such potential future changes requires significant effort involving both the engineers who adjust the rules and the legal experts who confirm the changes. Second, there is a need for manually-labeled datasets. While  rule-based approaches do not typically require training, creating datasets is still required for validating the rules and drawing conclusions on their generalizability. In the  regulatory compliance context, even small datasets are expensive to create since analyzing regulated documents requires legal background. 
Third, the complexity of the rules grows with the complexity of the legal text in the regulated documents. The DPA text is typically long, convoluted, and attempts to simultaneously address multiple GDPR provisions. Thus, complex rules are necessary for checking the completeness of DPAs against GDPR.      

Drawing on Amaral \textit{et al.}'s work, we propose in this paper ten alternative solutions which are based on different enabling technologies, including traditional machine learning (ML) as well as deep learning (DL) and the recent NLP technologies featured by large language models (LLMs) 
and few-shot learning (FSL) frameworks. The goal of our analysis is to examine these technologies side-by-side and assess how they fare in a challenging domain such as the legal domain, focusing on a particular use case; i.e., checking the completeness of DPAs against GDPR. 
%
Our comparative analysis aims to reveal the capabilities of the different solutions in analyzing legal text and understanding the  domain terminology. Our analysis will further shed the light on data scarcity, an important factor to consider in general when developing AI-enabled automation and in particular in the regulatory compliance context. Our findings provide researchers and practitioners with insights about how technologies compare against one another and help them select the most suitable technology for a new use case. 

\sectopic{Contributions.} Concretely, our paper makes the following  contributions:

\noindent \textbf{(1)} We formally define the completeness checking as a text classification problem. Given a set of GDPR provisions and a set of DPA sentences, completeness checking can be achieved by classifying the sentences as relevant or not relevant to each provision. This text classification task can be achieved via binary classification (e.g., ``provision x'' versus ``not provision x'') as well as multi-class classification where each provision is considered as one class. We elaborate the problem definition in Section~\ref{subsec:problem_definition}.    

\noindent \textbf{(2)} We propose ten alternative solutions for checking the completeness of DPAs against GDPR. Our solutions design is driven by two main objectives. The first objective is to compare the performance of different enabling technologies in checking the completeness of DPAs. We achieve this objective through running extensive experiments in different settings to draw conclusions. 
The second objective is to experiment with FSL solutions which require a much smaller labeled dataset compared to conventional solutions. To this end, we propose four solutions that utilize recent LLMs, five solutions based on traditional ML, and one that uses FSL framework. We describe our solutions design in Section~\ref{subsec:solutions_design}.

\noindent \textbf{(3)} We empirically evaluate the alternative solutions on 163 real DPAs. To do that, we used the 50 DPAs presented by Amaral \textit{et al.}~\cite{amaral:22} and further curated a dataset of 113 DPAs covering various sectors including cloud services, data analytics, and accounting audits. The curatation of the dataset was performed by three annotators who had legal background.  The annotators collectively analyzed 31,185 sentences, of which only 3,387 sentences were labeled as satisfying at least one provision in GDPR. We provide more details about the curation process in Section~\ref{subsec:datacollection}. 
On a subset of 30 DPAs that we used exclusively for evaluating the alternative solutions, we observe that LLMs-based solutions yielded the best performance with BERT and RoBERTA in the lead. BERT outperformed the rest of the solutions in solving the binary classification problem with a precision and recall of 75.1\% and 90.1\%, respectively. RoBERTa achieved the best performance in solving the multi-class classification problem with a precision and recall of 69.8\% and 96.6\%. 

Our evaluation further shows that using FSL yields better precision on the cost of a significant loss in recall. Compared to BERT on the binary task, the FSL-based solution has an average gain in precision of $\approx$3 percentage points (pp) and an average loss in recall of $\approx$10 pp. Compared to RoBERTa on the multi-class classification task, the FSL-based solution achieves $\approx$11 pp more precision and $\approx$17 pp less recall. These results are expected since FSL is developed on a small proportion of our dataset (30\% of the entire dataset in our case). Moreover, FSL does not require specific hardware to develop and run. With that in mind, FSL can still be useful in some contexts.
We report on our empirical evaluation in Section~\ref{sec:evaluation}.


\noindent \textbf{(4)} We analyze the impact of the dataset imbalance and size, two major characteristics that can significantly affect the performance of the alternative solutions. Specifically, we experiment with various methods, including random sampling and data augmentation to improve the distribution of the underrepresented examples in our dataset. 
In brief, our experiments indicate that random oversampling performs better than data augmentation techniques with an average gain of about 4 pp in accuracy.  We also show that the best strategy to balance the dataset is to use a combination of techniques to increase the minority class and simultaneously decrease the majority class. This strategy has been already applied in the RE literature~\cite{Rasiman:22}. 
Further details are given in Section~\ref{sec:rqs}.  

\sectopic{Structure.} The remainder of the paper is structured as follows: Section~\ref{sec:background-related} explains the necessary background for our alternative solutions and further reviews related work. Section~\ref{sec:automated_complaince_checking} defines the completeness checking problem and discusses our solutions design. Section~\ref{sec:evaluation} presents the details of our empirical evaluation  
and reports on the performance of the alternative solutions against one another and further provides insights on the impact of the size and imbalance in the dataset on their performance. 
Section~\ref{sec:threats} discusses threats to validity. Finally, Section~\ref{sec:conclusion} concludes the paper.

\section{Background and Related Work}
\label{sec:background-related}

In this section, we discuss the necessary background and further review the existing literature relevant to DPA completeness checking. 

\subsection{Background}
\label{sec:background}
Below, we briefly present GDPR, and then explain the different enabling technologies underlying our proposed solutions.

\sectopic{GDPR. }
\label{subsec:GDPR}
The GDPR~\cite{EU2018} is the European benchmark for data protection and privacy standards. According to GDPR, an organization can be a data controller or data processor. The controller decides the purpose and mechanism of collecting and processing personal data, whereas the processor processes personal data on behalf of the data controller. 
In our work, we focus on data processing agreements (DPAs). A DPA is a legally-binding contract between a data controller and data processor to ensure the protection of personal data throughout the data processing chain. A DPA sets out the rights and obligations of the controller and processor. To be deemed complete, the DPA's content should comply with the provisions  of Article 28 in GDPR.

\sectopic{Large Language Models (LLMs). }
LLMs, which have dominated the current state-of-the-art in NLP, are deep learning models that are pre-trained with a huge amount of (textual) data with the main objective to predict the next word in a text sequence. The pre-trained models obtain a basis knowledge about the languages to which they are exposed. These models can then be used to solve NLP downstream tasks such as text classification and question answering. There are two strategies for using the pre-trained LLMs, namely \textit{fine-tuning} the pre-trained models with task-specific labeled datasets, and \textit{extracting dense representations} (also called embeddings) that describe text sequences in the downstream task and then use these embeddings as learning features in an ML-based solution. Below, we briefly introduce the LLMs that we apply in our work.

\noindent (1) \textit{Bidirectional Encoder Representations from Transformers (BERT)~\cite{Devlin:18}} is one of the early LLMs which is still applied in the NLP literature due to its robust performance. BERT is pre-trained using two tasks: (1) predicting a randomly masked word in a text sequence by learning about its surrounding context to the left and right, and (2) predicting whether two sentences are consecutive. BERT is built using the Transformer architecture which is composed of a multi-layer encoder-decoder structure, wherein the encoder maps an input text sequence into numerical vector representations and the decoder generates an output sequence. BERT has 110 million parameters, and is pre-trained on 16GB of text corresponding to 3.3 billion words from the BooksCorpus~\cite{Zhu:15} and the entire English Wikipedia.

\noindent (2) \textit{A Lite BERT (ALBERT)~\cite{lan:19}} is a light variant of BERT. While ALBERT uses the same vocabulary size as BERT, it applies, unlike BERT, reduction techniques to reduce the number of parameters to a total of 12 million (instead of 110 million), increasing thereby the efficiency of the pre-training phase. ALBERT achieves comparable performance to BERT on different NLP benchmarks.

\noindent (3) \textit{Legal-BERT~\cite{Chalkidis:20}} is another variant of BERT that is pre-trained on 12GB of a legal text corpus crawled from publicly available web portals. The corpus contains the EU as well as the United Kingdom legislations, cases from the European Court of Justice and the European Court of Human Rights, and court cases and contracts from the United States (US). Legal-BERT outperforms BERT on two NLP tasks specific to the legal domain, namely multi-label text classification of EU laws and named entity recognition on US contracts.

\noindent (4) \textit{Robustly Optimized BERT pre-training Approach (RoBERTa)~\cite{Liu:19}} is a variant of  BERT with the main difference that it is pre-trained with a much larger dataset containing 161GB of text coming from CC-NEWS~\cite{nagel:16}, openWebText~\cite{gokaslan:19}, and Stories dataset~\cite{Zhu:15}. Additional differences compared to BERT include: (i) RoBERTa is trained only on one task, namely predicting the masked words; (ii) masking the words is done dynamically where different words are masked in different epochs; and (iii) the pre-training is performed for longer time on longer text sequences.

\noindent (5) \textit{Sentence-BERT} (SBERT) is a variant of BERT that is exposed to a large corpus of natural language inference, improving thereby the model's capability in deriving semantic representations of entire sentences instead of merely single tokens.  SBERT is the basis for the current state-of-the-art on LLMs for generating sentences embeddings.

In our work, we consider the above-explained strategies for utilizing LLMs. Consequently, we fine-tune the first four LLMs to perform the completeness checking of DPAs, whereas we use SBERT only for extracting the embeddings which then serve as the learning features in our ML-based alternative solutions. 

\sectopic{Few-shot Learning (FSL).}  With the rise of pre-trained LLMs, it became possible to train text classifiers using few  examples~\cite{pushp2017,schick2022}. FSL leverages the extensive knowledge that these pre-trained models have obtained being exposed to a huge body of unlabeled text. The few labeled examples are then used to teach LLMs about the specific classification task. 
The idea has been investigated in the RE literature for classifying textual requirements~\cite{alhoshan2023}. 
More recently, Tunstall \textit{et al.}~\cite{tunstall_setfit:22} propose SetFit, a framework for few-shot fine-tuning of sentence transformers. This framework is composed of two phases. In the first phase, it builds on SBERT model to learn sentences representations within the scope of the specific task, and for that it uses a handful of labeled examples. In the second phase, the sentences representations along with their labels are fed in to a classification head, resulting in a text classifier. This framework has shown promising results~\cite{bashir:23,halterman:23,halterman:23a,ali:23,kashyap:23}. In our work, we experiment with both ideas: fine-tuning pre-trained LLMs on a small set of labeled data as well as using SetFit framework.

\sectopic{Machine Learning (ML). } In our work, we use supervised ML algorithms for performing text classification. 
Being exposed to manually-labeled datasets of input examples mapped to a set of predefined output labels (or classes), such algorithms can learn to predict the most likely output class for a new, previously unseen input example. For checking the completeness of DPAs (the focus of our paper), an input example represents a text sequence in a given DPA and the output classes are one or more provisions in GDPR with which the DPA text is complying. Positive examples are those text sequences that satisfy the GDPR provisions, and negative examples are those that do not. 
As a pre-requisite for building ML classifiers, the input text must be converted to numerical representations. Current state-of-the-art represent text using embeddings, which are dense vector representations generated by LLMs. In our work, we represent the text using sentence embeddings from SBERT (described above).       
For our comparative analysis, we selected three well-known traditional ML algorithms, including \textit{logistic regression (LR)}, \textit{random forest (RF)}, \textit{support vector machine (SVM)}, and two deep learning (DL) algorithms, namely \textit{a multilayer perceptron (MLP)} ---a feedforward neural network, and \textit{bidirectional long short-term memory (BiLSTM)} ---a recurrent neural network. BiLSTM has been widely applied in the NLP literature prior to LLMs.

\sectopic{Dataset Handling. }
One of the major challenges encountered when developing automated solutions based on supervised learning, our work not being an exception, is the imbalance of the labeled dataset. In our dataset, the positive examples are significantly under-represented. To improve the distribution of the positive examples, we experiment with data augmentation techniques. To handle data imbalance, we further experiment with random sampling. Below, we introduce the 
techniques applied in our work. 

\noindent (1) \textit{Data Augmentation:} 
Data augmentation is used to automatically generate more training data (positive examples in our case). Data augmentation methods can be broadly categorized into~\cite{Li_DA:22}: (i) \textit{paraphrasing-based methods} where the original training examples are reproduced using semantically similar text, and (ii) \textit{noise-based methods} where the original examples are altered without affecting the meaning. 
In our evaluation, we use three paraphrasing-based methods, including \textit{back translation (BT)}, \textit{synonym replacement (SR)}, and \textit{embeddings replacement (ER)}. In BT, the original text in the training example is first translated from its language (English in our case) to other languages (e.g., French or German) and is then translated back to the original language, obtaining thereby a paraphrased version of the original text. In SR, one or more words in the training example are replaced by their synonyms using some external knowledge resource such as WordNet~\cite{Miller:95,Fellbaum:98}. In ER, randomly-selected words in the original training example are replaced with the other words which have the most similar embeddings in the vector space as generated by some pre-trained LLMs. In our work, we apply this method using the embeddings of three LLMs, namely ALBERT, BERT, and RoBERTa. 

We further use a \textit{noise injection (NI)} method which introduces random noise to the text in four ways~\cite{Li_DA:22}, including \textit{swapping} which swaps two randomly-selected words in a text sequence, \textit{deletion} which deletes random words in a text sequence, \textit{substitution} which substitutes random words in a text sequence, and \textit{cropping} which cuts off a random consecutive set of words in a text sequence.

\noindent (2) \textit{Data Imbalance Handling:}
To obtain a more balanced dataset,  random sampling can be used~\cite{branco:16}. In our evaluation, we apply  (i) \textit{random undersampling (RU)} where instances from the majority classes are randomly removed, (ii) \textit{random oversampling (RO)}, where instances for the minority classes are duplicated, and (iii) a mix of RU and RO where both the majority class is reduced and simultaneously the minority class is increased. 
\subsection{Related Work} 
\label{sec:related}

Completeness in RE has been extensively studied as part of quality assurance of requirements. A recent mapping study shows that requirements completeness is the second top most investigated quality attribute after ambiguity~\cite{montgomery:22}. Most of the existing approaches rely on external knowledge resources for detecting missing requirements. Ferrari \textit{et al.}~\cite{ferrari2014} uses NLP methods to identify missing concepts and relations in textual requirements according to available documents created during requirements elicitation phase (e.g., customer-meeting transcript). 
Arora \textit{et al.}~\cite{arora2019} verifies the completeness of a set of natural-language requirements against a domain model that is created a priori. More recently, Luitel \textit{et al.}~\cite{luitel:23} propose leveraging  LLMs such as BERT as an alternative external knowledge resource to identify incompleteness violations in requirements. 
In the context of regulatory compliance, the external resource is the regulation or an abstract representation thereof. Much work has been done so far for eliciting legal requirements, creating abstract representations and facilitating the navigation through regulations mainly to enable developing compliance software systems and/or checking the compliance of software against applicable laws. Existing work relies primarily on modeling the regulation~\cite{Torre:19,Pullonen:19,Zeni:15,Soltana:14,Ingolfo:14,otto:07}. There are also attempts to retrieve compliance-relevant information by querying the regulation~\cite{Abualhaija:22,sleimi:19}. Dealing with data scarcity is another relevant research topic. Gebauer \textit{et al.}~\cite{gebauer:23} propose integrating ML into the manual annotation process in privacy policies, focusing on data subject rights. The authors employ LLMs to devise a classifier that predicts based on previously seen training examples whether unseen text is about data subject rights. The predictions are leveraged to help human annotators annotate unseen privacy policies more efficiently.    

Another research strand in RE focuses on the completeness of regulated documents against applicable regulations. The intuition behind is that regulated documents are often more focused on certain aspects and are thus easier to navigate and elicit legal requirements. For example, DPAs set out the obligations of the processor. Bhatia \textit{et al.}~\cite{Bhatia:19} develop an automated approach for detecting the incompleteness violations in privacy policies against the privacy regulations. Their approach applies semantic frames where a sentence is analyzed according to what semantic roles certain verbs expect. For instance, the verb ``buy'' expects the roles of an actor (i.e., the buyer), something bought, and a price. This decomposition enables detecting incompleteness according to what roles are missing in the sentence. Torre \textit{et al.}~\cite{Torre:20RE} and Amaral \textit{et al.}~\cite{Amaral:21b} propose using a combination of NLP and ML technologies to detect the incompleteness violations in privacy policies. Specifically, they classify the textual content of a privacy policy according to a comprehensive conceptual model consisting of semantic metadata manually-defined based on relevant GDPR provisions. 

The closest to our work is the one by Amaral \textit{et al.}~\cite{amaral:22} who propose using semantic frames combined with rules to check the completeness of DPAs against GDPR. In collaboration with legal experts, they first define a set of 45 compliance criteria that are derived from GDPR provisions concerning DPAs. These criteria are documented as ``shall''-requirements to be better understood by requirements engineers. Out of the total 45 criteria, 26 are mandatory, meaning that the DPA's content must satisfy these criteria lest the DPA is non-compliant. The mandatory criteria are further broken down into three categories, namely metadata, processor obligations, and controller rights. Using rules on top of semantic frames, they develop then an automated approach to verify the content of DPAs according to these criteria. 

Similar to the existing literature in RE, our work focuses on verifying the completeness of a regulated document (a DPA, in our case) against the applicable regulations (the provisions of GDPR). 
In contrast with the existing literature, we (1) conduct a comparative analysis of ten alternative solutions covering different enabling technologies, and further (2) provide insights about the impact of the size and imbalance of the training data on these solutions. 
Specifically, our work builds on the legal knowledge concerning the interpretation of GDPR with regard to DPAs completeness as established by Amaral \textit{et al.}~\cite{amaral:22}. We scope our work to 19 mandatory compliance criteria concerning processor obligation, listed in Table~\ref{tab:mandatory-requirements}. The motivation behind our choice is two-fold. First, these criteria are mandatory, i.e., they must be satisfied in the DPA, consequently in the software systems that are employed for data processing. Second, unlike other mandatory criteria which mainly constrain the binding agreement, e.g., its duration, criteria on processor obligation can be translated to actionable points in the software systems which process personal data. For example, a criterion stating that the processor must ensure the security of personal data indicate the necessity for functional requirements like the software shall apply cryptography when transferring personal data.           

\begin{table}
\caption{GDPR Provisions concerning Mandatory Processor's Obligation in DPAs~\cite{amaral:22}.}
\label{tab:mandatory-requirements}
 \footnotesize
  \centering
   \begin{threeparttable}[t]
  \begin{tabularx}{0.98\textwidth} {@{} p{0.06\textwidth}  
  @{\hskip 0.5em}  X}
   \toprule
    ID  & Provision \\ 
    \midrule
    \textbf{P1} & The processor shall not engage a sub-processor without a prior specific or general written authorization of the controller. \\ 
    \midrule
    \textbf{P2} & In case of general written authorization, the processor shall inform the controller of any intended changes concerning the addition or replacement of sub-processors.  \\
    \midrule
    \textbf{P3} & The processor shall process personal data only on documented instructions from the controller. \\
    \midrule
    \textbf{P4} & If the processor requires by Union or Member State law to process personal data without instructions and law does not prohibit informing the controller on grounds of public interest, the processor shall inform the controller of that legal requirement before processing. \\
    \midrule
    \textbf{P5} & The processor shall ensure that persons authorized to process personal data have committed themselves to confidentiality or are under an appropriate statutory obligation of confidentiality. \\
    \midrule
    \textbf{P6} & The processor shall take all measures required pursuant to Article 32 or to ensure the security of processing. \\ 
    \midrule
    \textbf{P7} & The processor shall assist the controller in fulfilling its obligation to respond to requests for exercising the data subject's rights. \\ 
  \midrule
  \textbf{P8} & The processor shall assist the controller in ensuring the security of processing. \\ 
  \midrule
  \textbf{P9} &  The processor shall assist the controller in notifying a personal data breach to the supervisory authority. \\ 
  \midrule
  \textbf{P10} &  The processor shall assist the controller in communicating a personal data breach to the data subject.  \\ 
  \midrule
  \textbf{P11} &  The processor shall assist the controller in ensuring compliance with the obligations pursuant to data protection impact assessment.  \\ 
  \midrule
  \textbf{P12} & The processor shall assist the controller in consulting the supervisory authorities prior to processing where the processing would result in a high risk in the absence of measures taken by the controller to mitigate the risk. \\
  \midrule
  \textbf{P13} & The processor shall return or delete all personal data to the controller after the end of the provision of services relating to processing. \\ 
  \midrule
  \textbf{P14} & The processor shall immediately inform the controller if an instruction infringes the GDPR or other data protection provisions. \\
  \midrule
  \textbf{P15} & The processor shall make available to the controller information necessary to demonstrate compliance with the obligations Article 28 in GDPR.  \\
  \midrule
  \textbf{P16} & The processor shall allow for and contribute to audits, including inspections, conducted by the controller or another auditor mandated by the controller. \\ 
  \midrule
  \textbf{P17} & The processor shall impose the same obligations on the engaged sub-processors by way of contract or other legal act under Union or Member State law. \\
  \midrule
  \textbf{P18} & The processor shall remain fully liable to the controller for the performance of sub-processor's obligations. \\
  \midrule
  \textbf{P19} & When assessing the level of security, the processor shall take into account the risk of accidental or unlawful destruction, loss, alternation, unauthorized disclosure of or access to the personal data transmitted, stored or processed. \\
   \bottomrule
 \end{tabularx}
    \end{threeparttable}
\end{table}

\section{Automated Completeness Checking of DPAs}
\label{sec:automated_complaince_checking}
In this section, we first provide a generic definition of the completeness checking against a regulation as a text classification problem.  
We then explain the design of our proposed alternative solutions.

\subsection{Problem Definition}
\label{subsec:problem_definition}
\begin{definition}[Completeness Checking]
Let $\mathcal{S} = (s_1, s_2, \ldots, s_n); n\geq 1$ be a list of text span in a given regulated document (denoted as $\mathcal{D}$), where $n$ is the number of sentences. Let $\mathcal{P} = (p_1, p_2, \ldots, p_m); m\geq 1$ be a set of  provisions in some regulation (denoted as $\mathcal{R}$) related to the completeness of the regulated document, where $m$ is the number of provisions. The completeness checking can be defined as a text classification problem as follows:

\begin{itemize}
    \item \textit{Binary classification:} For each provision $p_i; 1\leq i\leq m$, we build a corresponding binary classifier $c_i; 1\leq i\leq m$ that predicts whether a given text span $s_j; 1\leq j\leq n$ \textit{satisfies} $p_i$. This definition allows multi-labeling, i.e., the same span can satisfy multiple provisions. 
    \item \textit{Multi-class classification:} For the set of provisions, $\mathcal{P}$, we build one classifier $c_p$ which predicts whether a given text span $s_j; 1\leq j\leq n$ \textit{satisfies} any  provision in $\mathcal{P}$. This definition is restricted to single-labeling, i.e., a span can satisfy exactly one provision.
\end{itemize}

Following this, $\mathcal{D}$ is complete according to $\mathcal{R}$ when there is, for each provision $p_i$, at least one text span $s_j$ in $\mathcal{D}$ which satisfies $p_j$. Otherwise, $\mathcal{D}$ is considered as incomplete. A provision that is not satisfied in $\mathcal{D}$ leads to an \textit{incompleteness violation}.
\end{definition}

Applying this definition on our work, $\mathcal{R}$ represents GDPR and $\mathcal{D}$ represents a DPA. We consider 19 provisions from GDPR, i.e., $m=19$. In our experiments, we build both binary classifiers as well as multi-class classifiers as we discuss later in this section. The overall goal of the alternative solutions in our work is to reveal the incompleteness violations in a given DPA, i.e., the GDPR provisions which the DPA does not satisfy.

\subsection{Alternative Solutions Design}
\label{subsec:solutions_design}
\begin{figure} [t!]
\centering
\includegraphics[width=0.98\textwidth]{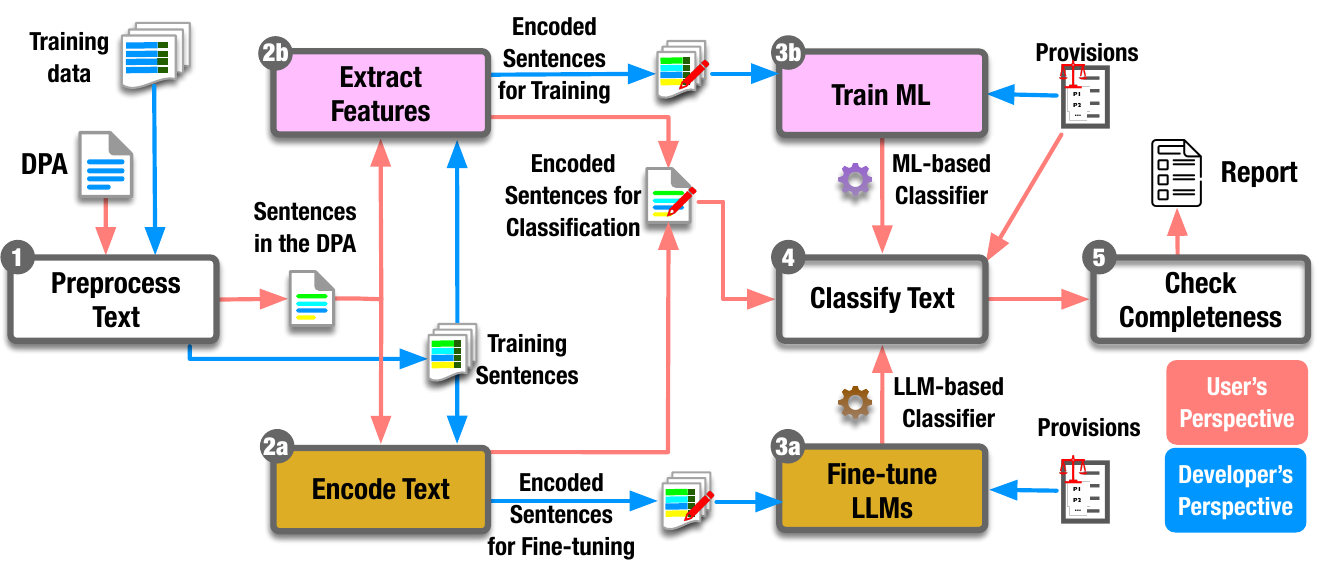}
\caption{Overview of our Solution Design.}
\label{fig:approach}
\end{figure}

In this section we discuss the design of our proposed solutions for checking the completeness of a DPA against GDPR. The solutions can be categorized into (A) LLM-based which also covers the FSL solution, and (B) ML-based which spans several traditional and recent ML algorithms. 
The main difference between the two categories is that LLM-based solutions are pre-trained on a massive amount of text, i.e., they have obtained knowledge about the language We fine-tune the pre-trained LLMs on completeness checking utilizing thereby transfer learning concepts.
In contrast, ML-based solutions are directly trained on the specific task at hand without having any prior knowledge. In Section~\ref{sec:evaluation}, we investigate the benefit of transfer learning. 
With the intuition to keep our solutions design versatile, i.e., adjustable to more recent alternative algorithms in each category, we describe below the different steps in each category without referring to specific algorithms. 

The overview of our solutions is depicted in Fig.~\ref{fig:approach}. 
Note that the figure distinguishes between the user's and developer's perspectives. The user's perspective assumes that the classifiers used in Step~4 are readily available, whereas the developer's perspective needs to build these classifiers. 
Step~1 preprocesses the textual content of a single DPA or a training set of several DPAs. Step~2 represents the input text as feature vectors compatible with the intended solution. To do that, Step~2(a) encodes the text, and  Step~2(b) extracts a set of feature embeddings following the common practices in the current ML literature. 
Based on the results of Step~2, Step~3(a) fine-tunes an LLM-based and Step~3(b) trains an ML-based solution over the training dataset. Step~4 predicts whether the DPA's text satisfies the GDPR provisions. 
Finally, Step~5 post-processes the predictions to conclude the violations in the DPA producing thereby a detailed report as the final output. Below, we elaborate these steps. 

\subsubsection{Step~1: Preprocess Text}
The input to this step can be either a single DPA or a set of several DPAs used for creating the solutions. In both cases, we parse each DPA and preprocess its text using a simple NLP pipeline, composed of the three modules: The first module is a \textit{tokenizer} which splits the text into a set of tokens, e.g., words and punctuation marks. The second module represents a \textit{sentence splitter} which splits the text into sentences according to widely-used endings of a sentence captured in the punctuation marks. Examples of such endings include periods and colons in the English language. We note that the sentence splitter does not necessarily produce grammatically correct sentences. Finally, the third module is a \textit{normalizer} that replaces the specific names and references of the processors and controllers by their generic descriptions, i.e., \textit{PROCESSOR} and \textit{CONTROLLER}. While the first two modules are fully automated, the third module is semi-automated using regular expressions and human in the loop. The reason for semi-automating is due to the multiple ways of referring to the processors and controllers which vary across the DPAs in our dataset. For instance, a processor can be referred to by its name (i.e., ORGANIZATION X) or using different alternative references including \textit{importer}, \textit{service provider}, etc. 
The intermediary output of this step represents the list of sentences in the DPA, i.e., $\mathcal{S} = (s_1, s_2, \ldots, s_n); n\geq 1$ where $n$ is the number of sentences in the DPA. This output is passed on to Step~2. 

\subsubsection{Step~2: Create Feature Vectors} 
In this step, we loop over the sentences in the input DPA and create for each sentence a corresponding feature vector which will be used to enable building the solutions or classifying the text. Below, we discuss this step considering the two enabling technologies considered in our work.
\sectopic{(a) Encode Text for LLM-based Solutions. }
LLMs require the input text to be encoded in a specific format. BERT-like models automatically add two special tokens, namely \textit{[CLS]} and \textit{[SEP]}. The [CLS] encapsulates the text representation which is fed into an output layer for classification. The [SEP] token is used to split two input text pairs which not applicable in our case as the input is a single sentence. Given a sentence $s_j \in \mathcal{S}$, where $s_j$ is a sequence of tokens denoted as $t_1, \ldots, t_k; k\geq 1$, this step encodes $s_j$ as $([CLS],t_1,\ldots,t_k[SEP]\phi)$. This encoded text is passed on to Step~3(a) and Step~4 for fine-tuning LLM-based solution or applying readily-available ones to classify the text according to the provisions, respectively.

\sectopic{(b) Extract Features for ML-based Solutions. }
As discussed in Section~\ref{sec:background}, text classification using ML requires transforming the text into mathematical representations which are then used as learning features. In our work, we use the most recent representation methods to generate sentences embeddings (denoted as $e$). A sentence $s_j$ will be represented by a sequence of numbers corresponding to its embeddings $e_1, \ldots, e_{768}$. Note that the sentences embeddings are 768-dimensional since we apply SBERT (a variant of BERT) to generate these embeddings. 
Similar to Step~2(a), the intermediary output of this step is passed on to Step~3(b) to train the ML classifiers or Step~4 to classify text.

\subsubsection{Step~3: Build Text Classifiers} 
This step is applied only in the developer's perspective. The goal is to create different classification models that can be then used for classifying the text of a given DPA according to whether they satisfy the GDPR provisions. For creating the classifiers, we use a training dataset (elaborated in Section~\ref{subsec:datacollection}) which is subjected to the first two steps, namely preprocessing and feature extraction. 
Referring to the problem definition described earlier, we build binary and multi-class classifiers as follows.  For a particular provision $p_i$, a corresponding binary classifier is created by being exposed to positive examples, i.e., all sentences in the training set that satisfy $p_i$, and negative examples, i.e., all sentences in the training set (or a subset thereof) that do not satisfy $p_i$. As a result, we created a total of 19 binary classifiers corresponding to the GDPR provisions considered in our work. 
We further create one multi-class classifier that is exposed to all sentences satisfying any GDPR provision. 
Following this, we \textit{(a) fine-tune LLMs}, namely ALBERT, BERT, Legal-BERT, and RoBERTa, and \textit{(b) train five ML algorithms}, namely LR, RF, SVM, MLP, and BiLSTM. See Section~\ref{sec:background} for more details on these models. The hypeparameters of all models are optimized on a subset of the training set.

\subsubsection{Step~4: Classify Text}
This step takes as input the provisions of GDPR, a classification model (created in Step~3) and the representation of a sentence in the input DPA. Each model then predicts whether the sentence satisfies a provision or not. In a realistic scenario, the user would use only one solution and one problem definition. For instance, the user will apply one multi-class classifier to predict the provision discussed in the given sentence. To help the user decide what is more suitable for their application context, we experiment in Section~\ref{sec:evaluation} with multiple alternative solutions and report on their performance.

\subsubsection{Step~5: Check Completeness}
Finally, Step~5 takes the predictions made for all the sentences in the input DPA and process them to predict GDPR-related completeness violations as follows: A provision $p_i$ is violated in a given DPA if there are no sentences that are predicted to be on $p_i$ in the DPA. Otherwise, if there is at least one sentence on $p_i$ in the DPA, then $p_i$ is satisfied. Any missing $p_i$ leads to an incomplete DPA according to GDPR. The final output of this step is then a detailed report describing the missing provisions as well as the satisfied ones alongside the sentences that actually satisfy them.

\section{Empirical Evaluation} 
\label{sec:evaluation}
In this section, we evaluate our proposed solutions.

\subsection{Research Questions}
Our empirical evaluation addresses the following research questions (RQs): 

\sectopic{RQ1: Which alternative solution is the most accurate for checking the completeness of DPAs against GDPR?} \\
RQ1 thoroughly compares our proposed solutions aiming at determining the most accurate one for completeness checking of DPAs. We analyze the performance of the solutions in solving the completeness checking both as a binary and multi-class classification problem. The best-performing solution is then used to address the subsequent RQs.

\sectopic{RQ2: Which data imbalance handling method yields the best accuracy for DPA completeness checking against GDPR?} \\
As we discuss in Section~\ref{subsec:datacollection}, our dataset is highly skewed towards the negative examples (i.e., sentences that do not satisfy any provision) being significantly more with the positive examples (i.e., DPA sentences that satisfy any GDPR provision).   RQ2 investigates several methods that handle data imbalance, and further analyzes the impact of applying these methods on the results of the alternative solutions. 

\sectopic{RQ3: How accurate are FSL solutions in checking the completeness of DPAs against GDPR?}
Scarcity of labeled datasets is a problem that has been often discussed in the RE literature. In the context of regulatory compliance, building labeled datasets often requires the involvement of legal experts which is expensive and not always possible. To address this concern, we apply an FSL framework which utilizes LLMs, yet requires much less training data. In RQ4, we assess using FSL in several scenarios and discuss its performance compared with the other alternative solutions developed over the entire training dataset. 

\sectopic{RQ4: What is the execution time of our proposed solutions?} \\
Our solutions are intended to assist requirements engineers in identifying incompleteness violations in DPAs which can significantly impact the compliance requirements specified for data processing activities in software. Our solutions can also assist legal experts in their compliance checking activities. To run any solution in a realistic scenario, it must have practical time. 
RQ4 investigates the execution time of our proposed solutions. 

\subsection{Implementation Details and Data Availability}
We implemented all alternatives solutions using Python 3.8~\cite{guido_python:09}, in PyCharm IDE~\cite{pycharm:15}. 
For operationalizing the NLP pipeline applied in Step~1, we use the NLTK 3.7 library~\cite{bird_nltk:06}. In Step~2(b), we extract the feature embeddings from SBERT (concretely, `all-MiniLM-L6-v2') which is available in the Transformer 4.18.0 library~\cite{transformers}. 
We use the same Transformer library and PyTorch 1.12.1~\cite{paszke_pytorch:19} for implementing MLP and BiLSTM in addition to all details concerning the LLM-based solutions, including the pre-trained models, the fine-tuning process, and the text encoding. All LLMs used in our study are the pre-trained \textit{BASE} variants, including `bert-base-uncased', `nlpaueb/legal-bert-base-uncased', `roberta-base', and `albert-base-v2'.  
We implement the FSL framework using the SetFit 0.6.0 library~\cite{tunstall_setfit:22} and sentence transformers 2.2.2 library~\cite{Reimers:19}. Finally, we use the Scikit-learn 1.0.2 library~\cite{pedregosa_scikit:11} which provides the basic functions for ML, e.g., training ML classifier, dataset imbalance handling and splitting the dataset. We make all our non-proprietary material used in our empirical evaluation publicly available at this link\footnote{\href{https://figshare.com/s/77338e558ffb6adf6f55}{https://figshare.com/s/77338e558ffb6adf6f55}}.

\subsection{Data Collection Procedure}
\label{subsec:datacollection}
The purpose of our data collection is to collect a large set of DPAs that are manually analyzed and checked for completeness against GDPR. To do so, we extended the original dataset (consisting of 50 DPAs) that is presented by Amaral \textit{et al.}~\cite{amaral:22} with another 113 DPAs, of which 17 are from diverse public resources. The DPAs in our dataset are contracts between controllers and processors coming from different sectors, such as tax and cloud services, services related to human resources, data analytics, accounting, audits, and pension services. 

To build the ground truth in our work, the DPAs were manually analyzed by three third-party annotators (non-authors). All annotators were in the graduate program at law department. Thus, they had the necessary legal background for our task. The annotators further attended a half-day training about compliance and completeness against GDPR. 
In this training, we introduced the GDPR provisions (in Table~\ref{tab:mandatory-requirements}) concerning processor's obligations. We prepared the annotation process as follows: We first take out 10\% of the DPAs to be used as a shared subset on which the interrater agreement is measured. We then split the DPAs into three subsets maintaining to the possible extent nearly equal number of sentences in each subset to balance the load. 
We provided clear instructions for analyzing the DPAs, including to examine each sentence in the DPA and decide whether the sentence satisfies one or more GDPR provisions. To ensure that the annotators obtained mutual understanding of the task, we let all annotators start by annotating the 10\% overlapping subset in such a way that each DPA is analyzed by at least two annotators. We then computed Cohen’s Kappa interrater agreement metric ($\kappa$)~\cite{Cohen:60}. The average $\kappa$ was $0.82$, implying ``almost perfect agreement''~\cite{Viera:05}. 
To solve the disagreements, we presented to each annotator the sentences which are disagreed upon by the other annotator on the same set, and asked them to provide feedback on whether they agree to change their annotation accordingly. All remaining disagreements were discussed in an online session. Following this, each annotator was assigned a subset of DPAs to analyze.

To mitigate fatigue, the annotators were recommended to work two hours at a time. They declared a total of $\approx$57 hours each over a span of three months. 

Our data collection procedure resulted in analyzing a total of 169 DPAs consisting of 31,185 sentences. Of these sentences, only a small fraction of $\approx$12\% (about 3,387 sentences) was labeled as satisfying any provision from GDPR. These sentences together with their labels are then used as our ground truth. 
We split the DPAs in our ground truth into two proportions as follows: 70\% of the DPAs is used for developing the alternative solutions and the remaining 30\% is used for assessing the performance of the solutions. The training and evaluation sets are thereafter referred to as $\tau$ and $\epsilon$, respectively. 
Table~\ref{tab:stats} provides the statistics about the total number of sentences as well as the total number of DPAs in each set that satisfy each GDPR provision. 

\begin{table*}[!t]
\caption{Results of our Data Collection Procedure.} \label{tab:stats}
    \begin{tabularx}{\textwidth}{@{}*{7}{>{\centering\arraybackslash}X}@{}}
    \toprule
    \multirow{2}{*}{Provisions} & \multicolumn{3}{c}{No. of DPAs} & \multicolumn{3}{c}{No. of Sentences} \\
    \cmidrule(lr){2-4}\cmidrule(lr){5-7}
    & $\tau$ & $\epsilon$ & $\sum$ & $\tau$ & $\epsilon$ & $\sum$ \\
    \midrule
    P1 & 120 & 19 & 139 & 234 & 40 & 274\\
    P2 & 68  & 17 & 85 & 83 & 21 & 104\\
    P3 & 113 & 28 & 141 & 168 & 50 & 218\\
    P4 & 58  & 12 & 70 & 73 & 20 & 93\\
    P5 & 110 & 24 & 134 & 155 & 34 & 189\\
    P6 & 76 & 17 & 93 & 627 & 100 & 727\\
    P7 & 115 & 24 & 139 & 244 & 41 & 281\\
    P8 & 28 & 1 & 29  & 30 & 1 & 31\\
    P9 & 46 & 6 & 52 & 57 & 7 & 64\\
    P10 & 16 & 1 & 17 & 18 & 1 & 19\\
    P11 & 68 & 19 & 87 & 88 & 20 & 108\\
    P12 & 28 & 2 & 30 & 30 & 2 & 32\\
    P13 & 127 & 25 & 152 & 202 & 38 & 240\\
    P14 & 79  & 15 & 94 & 89 & 16 & 105\\
    P15 & 86 & 13 & 99 & 165 & 22 & 187\\
    P16 & 98 & 20 & 118 & 192 & 33 & 225\\
    P17 & 113 & 22 & 135 & 205 & 31 & 236\\
    P18 & 97 & 17 & 114 & 127 & 26 & 153\\
    P19 & 64  & 12 & 76 & 84 & 13 & 97\\ 
\bottomrule
\end{tabularx}
\end{table*}

\subsection{Evaluation Procedure}
\label{subsec:eval_procedure}
To answer the RQs, we conduct the following experiments (EXPI -- EXPIV), explained below. 

\sectopic{EXPI. } 
EXPI answers RQ1. Specifically, we assess the performance of our proposed solutions both for performing binary and multi-class text classification to check the completeness of a given DPA against GDPR. The solutions are evaluated exclusively on $\epsilon$ (our evaluation set). To evaluate the results, we define for a given provision $p_j$, \textit{true positives} (TPs) as the cases for which $p_j$ is correctly found as satisfied in the DPA, \textit{false positives} (FPs) as the cases when $p_j$ is wrongly found to be satisfied, \textit{false negatives} (FNs) as the cases when a satisfied $p_j$ is missed by the solution (i.e., found as violated), and \textit{true negatives} (TNs) as the cases when $p_j$ is correctly found as not satisfied (i.e., violated) in the DPA. Following this, we apply the traditional metrics accuracy (A), precision (P) and recall (R), computed as A=(TP+TN)/(TP+TN+FP+FN), P=TP/(TP+FP), and R=TP/(TP+FN). In our work, recall is more important than precision. The intuition is that recall reflects the errors of how often the solution misses a satisfied provision (i.e., FNs). The solution missing any provision implies that the analyst will remain unaware of that provision in the DPA. From an RE standpoint, this causes incompleteness which can affect the overall compliance of the developed software. Precision on the other hand reflects the falsely predicting a satisfied provision in the DPA by the solution (i.e., FPs). Since such errors are associated with actual sentences found in the DPAs (which falsely indicate the provision), FPs (if not too many) can be more easily filtered out by the analyst.
Therefore, we further compute in EXPI $F_{\beta}$ score for $\beta=2$, indicating that recall is favored over precision. F$_2$ is computed as F$_2$ = $((2^2+1)*P*R$)/$(2^2*P+R)$.

\sectopic{EXPII. } EXPII answers RQ2. 
Specifically, we investigate which data imbalance handling strategy yields the most accurate solutions. 
Recall from Section~\ref{subsec:problem_definition} that we train 19 binary classifiers, one classifier for each provision $p_j \in \mathcal{P}$. In the training dataset $\tau$, the \textit{positive examples} include all sentences that satisfy $p_j$ in all DPAs in $\tau$. 
Conversely, the \textit{negative examples} include the sentences that do not satisfy $p_j$, i.e., the sentences that satisfy any provision but $p_j$ in addition to the sentences that do not satisfy any provision. In this case, the positive examples represent the minority classes, whereas the negative examples represent the majority. 
Table~\ref{tab:stats} reports the total number of sentences in $\tau$ satisfying each $p_j$, i.e., positive examples. 
We further train one multi-class classifier over the positive examples that satisfy the 19 provisions. In this case, we treat all sentences in $\tau$ that do not satisfy any provision as another class referred to as \textit{other}. The rationale behind including \textit{other} is that the overall accuracy of the multi-class classifier in distinguishing provisions improves due its the exposure to the text segments that do not satisfy any provision.  The multiple classes in our work include the 19 provisions (minority classes) and \textit{other} (majority class). 
In EXPII, we apply several data handling strategies aiming to balance $\tau$ through (1) increasing the minority classes, (2) decreasing the majority classes, and (3) a combination of both increasing and decreasing. 
We create 13 variants of $\tau$ using random sampling in combination with data augmentation (see Section~\ref{sec:background} for more details about these methods). 
Considering the best performing solutions resulted form EXPI, we redevelop these solutions by exposing the classifier each time to a different variant of $\tau$. We then apply the same metrics A, P, R, and F$_2$ as defined in EXPI to assess the performance of the resulting solutions over $\epsilon$.

\sectopic{Experiment III. } This experiment addresses RQ3. Specifically, we test the performance of the alternative solutions when they are developed on a much smaller proportion of the dataset equivalent to 30\%. In EXPIII, we also use SetFit framework  (explained in Section~\ref{sec:background}) to perform FSL with the same 30\% of the training dataset. All the different solutions are again tested exclusively on $\epsilon$. The purpose of EXPIII is to provide insights about how well the solutions fare when there are only few training examples available. To assess the performance we report on A, P, R, and F$_2$ as defined in EXPI. 

\sectopic{Experiment IV. }
This experiment addresses RQ4. We measure the running time of the different solutions taking into account the two perspectives described in Section~\ref{subsec:solutions_design}. In our work, we use a work station with AMD 12-Core processor (3.70GHz), INVIDIA GeForce RTX 3090 GPU, and 64GB of RAM for training and developing the classifiers, whereas we apply the developed classifiers to test the alternative solutions  on $\epsilon$ 
using a normal laptop with 8-Core Intel processor (2.4GHz), and 32GB of RAM.

\subsection{Answers to the RQs}
\label{sec:rqs}
\sectopic{RQ1: Which alternative solution is the most accurate for checking the completeness of DPAs against GDPR?} 

Table~\ref{tab:rq1.1} shows the accuracy, precision, recall, and F$_2$ of our proposed alternative solutions. The metrics are computed exclusively over $\epsilon$ to measure the capability of the solutions in identifying the satisfied provisions in the DPAs in $\epsilon$. The table shows the results considering the DPA completeness checking both as a binary classification and multi-class classification problem. 

\begin{table*}[!t]
\caption{Accuracy Results for DPA Completeness Checking Solutions  \textbf{(RQ1)}. } \label{tab:rq1.1}
  \footnotesize

   \begin{threeparttable}[t]
    \begin{tabularx}{\textwidth}{@{}l*{10}{>{\centering\arraybackslash}X}@{}}
    \toprule
    & \multicolumn{4}{c}{Binary Classification} & \multicolumn{4}{c}{Multi-class Classification} \\
    \cmidrule(lr){2-5}\cmidrule(lr){6-9}
    & A(\%) & P(\%) & R(\%) & F$_2$(\%) & A(\%) & P(\%) & R(\%) & F$_2$(\%) \\
    \midrule
    \ding{172} ALBERT & 69.1 & 82.4 & 51.0 & 55.2 & 78.1 & 79.6 & 77.2 & 77.7 \\
    \ding{173} BERT & \textbf{80.9} & 81.4 & \textbf{81.6} &  \colorbox{yellow!50}{\textbf{81.6}} & 83.0 & 82.7 & 84.7 & 84.3 \\
    \ding{174} Legal-BERT & 67.5 & 76.8 & 53.1 & 56.6 & 81.8 & 79.7 & 86.7 & 85.2 \\
    \ding{175} RoBERTa & 74.7 & 83.7 & 63.1 & 66.4 & \textbf{83.5} & 80.5 & \textbf{89.8} & \colorbox{yellow!50}{\textbf{87.8}} \\
    \cmidrule(lr){2-9}
    \ding{176} BiLSTM & \textbf{80.9} & 82.5 & 80.3 & 80.7 & 79.3 & 79.3 & 81.0 & 80.6 \\
    \ding{177} LR & 59.6 & \textbf{93.2} & 23.5 & 27.6 & 66.3 & \textbf{94.0} & 37.1 & 42.2 \\
    \ding{178} MLP & 77.5 & 85.8 & 67.7 & 70.7 & 78.8 & 86.5 & 69.7 & 72.5 \\
    \ding{179} RF & 65.6 & 87.1 & 39.1 & 44.0 & 63.9 & 83.0 & 39.8 & 44.4 \\
    \ding{180} SVM & 74.4 & 89.4 & 57.1 & 61.6 & 76.0 & 91.5 & 58.8 & 63.4 \\
    \bottomrule
\end{tabularx}
  \begin{tablenotes}
     \item[$\dag$]  The best values of A, P, R, and F$_2$ are highlighted in \textbf{bold}.   
     \end{tablenotes}
\end{threeparttable}
\end{table*}

With regard to binary classification, BERT achieves the best overall results in terms of accuracy, recall, and F$_2$. Compared to other LLMs, BERT achieves an average gain in accuracy of 10.5 percentage points ($pp$). BERT further achieves an average gain in accuracy of 11.6 $pp$ over all ML-based solutions except BiLSTM which yields the same accuracy as BERT. 
BERT achieves the best recall with a significant gain of 25.9 $pp$ and 34.7 $pp$ over LLM and ML-based solutions, respectively. Again, BiLSTM is an exception which performs on par with BERT and has a loss in recall of 1.3 $pp$.   
In contrast, we observe that ML-based solutions are performing better or on par with LLM-based solutions (in particular, BERT) in terms of precision. As shown in the table, LR yields the best precision value of 93.2\%, indicating a gain of 11.8 $pp$ over BERT. This is followed by the second best performing SVM with a precision of 89.4\%, RF with 87.1\%, and MLP with 85.8\%. BiLSTM achieves a comparable precision to the one achieved by LLM-based solutions. ML-based solutions (except BiLSTM) have very low recall ranging from 23.5\% for LR to 67.7\% for MLP which introduces a particularly notable degradation in the respective F$_2$ scores.   
Having a low recall indicates that many provisions were missed by these solutions. For instance, LR compared to BERT has an average loss in recall of $\approx$58 $pp$, entailing the risk of missing a large proportion of the satisfied provisions in $\epsilon$. On the one hand, the 11.8 $pp$ in precision (the advantage of LR over BERT) entails  a  fraction of falsely satisfied provisions introduced by BERT which can be filtered out by an expert with relative ease. The low recall of LR, on the othe hand, entails the risk of missing a lot of satisfied provision, compromising thereby the completeness of the DPA. Therefore, for binary classification, we select BERT as the best performing solution. 

With regard to multi-class classification, RoBERTa yields the best results in terms of accuracy, recall, and F$_2$. Compared to RoBERTa, legal-BERT and BERT achieve comparable performance in terms of F$_2$ with a loss of 2.6 $pp$ and 3.5 $pp$, respectively.
As shown in the table, ML-based solutions demonstrate similar behavior to the one observed for binary classification. That is, they achieve high precision values at the cost of low recall. BiLSTM is still an exception that achieves comparable results to LLMs. For multi-class classification, we select RoBERTa as the best performing solution.

The remarkably lower recall in the case of traditional ML-based solutions confirm that the text classification problem in the context of  completeness checking of DPA is complex. Not having any ground knowledge about the language (in contrast with LLM-based solutions) makes the prediction task for ML-based solutions more challenging. The reason could be due to the commonalities and textual overlaps between the different text segments in a given DPA which satisfy different provisions. BiLSTM, which was widely applied in the NLP literature prior to LLMs, performs exceptionally well. In addition to the fact that BiLSTM is built using recurrent neural networks, it is also effective since it learns the text in a bidirectional manner (similar to BERT). Bidirectional is essential in learning about the context and words co-occurrences in any language.   
LLM-based solutions obtained language capabilities during the pre-training phase and hence have an advantage over ML-based solutions (including BiLSTM) to better distinguish the DPA text.     

Overall we observe different behavior from the same LLM in binary versus multi-class classification. Recall that binary classifiers are created separately for each provision, e.g., the BERT-based model fine-tuned for predicting P1 is different from the one fine-tuned for P2. We believe that this can be the reason why the same model behaves differently. For instance, RoBERTa learned better in our context when it was exposed to all provisions at the same time (i.e., multi-class classification) whereas it learned poorly when presented with separate provisions against their negative examples.

\begin{table*}[!t]
\caption{Breakdown of F$_2$ Score per Provision for LLM-based Solutions. } 
\label{tab:rq1.2}
  \footnotesize

   \begin{threeparttable}[t]
\begin{tabularx}{\textwidth}{@{}*{10}{>{\centering\arraybackslash}X}@{}}
\toprule
&& \multicolumn{4}{c}{Binary Classification} & \multicolumn{4}{c}{Multi-class Classification} \\
\cmidrule(lr){3-6}\cmidrule(lr){7-10}
P& $|\epsilon|$ & \ding{172}  & \colorbox{yellow!50}{\ding{173}}  & \ding{174}  & \ding{175} & \ding{172}  & \ding{173} & \ding{174}  & \colorbox{yellow!50}{\ding{175}}  \\
\midrule
P1 & 19 & 89.5 & 94.7 & \textbf{95.7} & 90.4 & 86.0 & \textbf{88.5} & \textbf{88.5} & 85.1\\
P2 & 17 &39.5 & \textbf{77.4} & 0.0  & 0.0 & 67.1 & 68.8 & 76.5 & \textbf{78.3}\\
P3 & 28 &0.0  & \textbf{93.5} & 0.0  & 88.2 & 88.2 & \textbf{97.1} & 88.2 & 94.2\\
P4 & 12 & 82.0 & 88.7 & \textbf{89.6} & \textbf{89.6} & 78.1 & 85.9 & 88.7 & \textbf{95.2}\\
P5 & 24 & 96.6 & \textbf{98.4} & \textbf{98.4} & 88.2 & \textbf{95.8} & 95.0 & 95.0 & 95.0 \\
P6 & 17 & 0.0  & \textbf{61.8} & 0.0  & 0.0 & 46.0 & 59.5 & \textbf{73.0} & 71.4 \\
P7 & 24 & 84.0 & 87.5 & \textbf{90.9} & 87.5 & 80.5 & 84.0 & \textbf{97.6} & 94.3 \\
P8 & 1 & 0.0  & 0.0  & 0.0  & 0.0 & 0.0  & 0.0  & 0.0  & \textbf{41.7} \\
P9 & 6 & 0.0  & \textbf{46.9} & 0.0  & 32.3 & 57.1 & \textbf{62.5} & 27.8 & 46.9 \\
P10 & 1 & 0.0  & 0.0  & 0.0  & 0.0   
& 0.0  & 0.0  & 0.0  & 0.0 \\
P11 & 19 & 63.2 & \textbf{85.1} & 0.0  & 81.5 & 79.8 & 76.1 & \textbf{85.1} & \textbf{85.1} \\
P12 & 2 & 26.3 & \textbf{55.6} & 50.0 & 0.0 & 50.0 & 45.5  & \textbf{55.6} & \textbf{55.6} \\
P13 & 25 & 96.0 & 96.0 & 92.7 & \textbf{99.2} & 92.7 & \textbf{99.2} & \textbf{99.2} & \textbf{99.2} \\
P14 & 15  & 82.3 & 90.9 & \textbf{94.9} & 0.0 & 80.0 & 74.3 & \textbf{90.9} & \textbf{90.9}\\
P15 & 13 &  0.0  & 0.0  & 0.0  & 68.5 & 32.8 & 76.9 & 68.2 & \textbf{83.3} \\
P16 & 20 & 49.5 & 63.2 & \textbf{90.0} & 87.6 & 68.4 & 94.1 & 89.1 & \textbf{91.3} \\
P17 & 22 & 0.0  & \textbf{95.5} & 0.0  & 95.5 & 91.7 & 95.5 & 91.7 & \textbf{99.1} \\
P18 & 17 & 79.3 & \textbf{93.0} & 0.0  & 0.0 & 84.3 & 89.3 & \textbf{94.1} & \textbf{94.1}  \\
P19 & 12 & 28.3 & 67.2 & \textbf{69.2} & 0.0 &  \textbf{75.8} & 74.6 & 68.2 & 68.2  \\
\bottomrule
\end{tabularx}
  \begin{tablenotes}
 \item[$\dag$] P: Provision, $|\epsilon|$: The number of DPAs satisfying P in $\epsilon$. 
 \item[$\ddag$] \ding{172} ALBERT, \ding{173} BERT, \ding{174} Legal-BERT, \ding{175} RoBERTa.   
 
 \end{tablenotes}
\end{threeparttable}
\end{table*}
To better understand the behavior of LLMs on predicting the satisfied provisions in a DPA, we provide a breakdown of the F$_2$ scores per provision in Table~\ref{tab:rq1.2}. 
Considering the number of DPAs that contain the provision in the evaluation set ($|\epsilon|$), it is not a surprise that BERT scores zero in F$_2$ for P8 and P10 which are satisfied only in one DPA. We add to that that P8, P10, and P12 have the lowest number of positive examples in our training dataset as show in Table~\ref{tab:stats}. The difference between these provisions is that P12 has more context in GDPR (see Table~\ref{tab:mandatory-requirements}). We observed that the sentences describing P12 are as well longer than the ones describing P8 or P10, in which case helps the classifiers better learn how to distinguish P12 than even when P12 has few training examples.   
For binary classification, the table shows that BERT outperforms the other LLM-based solutions in nine provisions, namely P2, P3, P5, P6, P9, P11, P12, P17, and P18. Legal-BERT achieves good performance, and outperforms the other solutions in six provisions, namely P1, P4, P7, P14, P16, and P19. Given that the models are different for each provision, it is possible to create a hybrid classifier that combines the best performing classifiers based on BERT and Legal-BERT. Creating such a hybrid solution can slightly improve F$_2$ to 84.1\%, gaining thereby 2.5 \textit{pp} over selecting BERT. 

Unlike binary classifiers, the multi-class classifiers are developed on the entire set of provisions, i.e., we have one classifier per LLM-based solution. We observe more variation in the behavior of the best performing solutions than for binary classification. RoBERTa still performs the best in 11 out 19 provisions. Legal-BERT outperforms RoBERTA in three provisions (P1, P6, and P7) and scores equally good for five provisions (P11-P14 and P18). BERT outperforms the rest in two provisions (P3 and P9) and scores equally good in another two provisions (P1 and P13). Despite the overall worse performance, ALBERT still outperforms the rest in two provisions, namely P5 and P19. However, the average gain is not significant compared to the performance of the other models for P5 and BERT for P19.  The consistent good performance of RoBERTa confirms our decision to select it as the most accurate solution for multi-class classification.

\begin{tcolorbox}[arc=0mm,width=\columnwidth,
                  top=0mm,left=0mm,  right=0mm, bottom=0mm,
                  boxrule=1pt,colback=yellow!5!white,colframe=black]
\textbf{The answer to RQ1 is} LLM-based solutions outperform ML-based solutions in checking the completeness of DPAs against GDPR. When the completeness checking is formulated as a binary text classification problem, the most accurate solution is based on BERT with an average F$_2$ of 81.6\%. Alternatively, RoBERTa is the most accurate solution, yielding an F$_2$ of 87.8\%, when formulating the problem as a multi-class classification. 
\end{tcolorbox}

\sectopic{RQ2: Which data imbalance handling method yields the best accuracy for DPA completeness checking against GDPR?}

Table~\ref{tab:rq2} compares the accuracy results of the best performing solutions from RQ1, i.e., \ding{173} BERT for binary classification and \ding{175} RoBERTA for multi-class classification, when fine-tuned on 13 variants of our original training dataset. As explained in Section~\ref{subsec:eval_procedure}, these variants are generated through a combination of random sampling (\textit{RO} and \textit{RU}) and data augmentation methods (\textit{BT}, \textit{ER}, \textit{NI}, \textit{SR}). 

\begin{table*}[!t]
\caption{Accuracy Results for Data Handling Strategies  \textbf{(RQ2)}. } \label{tab:rq2}
  \footnotesize

    \begin{threeparttable}[t]
    \begin{tabularx}{\textwidth}{@{}p{0.1\textwidth}*{8}{>{\centering\arraybackslash}X}@{}}
    \toprule
    \multirow{2}{*}{Dataset$\dag$}& \multicolumn{4}{c}{Binary Classification (\colorbox{yellow!50}{\ding{173}})} & \multicolumn{4}{c}{Multi-class Classification (\colorbox{yellow!50}{\ding{175}})} \\
    \cmidrule(lr){2-5}\cmidrule(lr){6-9}
    & A(\%) & P(\%) & R(\%) & F$_2$(\%) & A(\%) & P(\%) & R(\%) & F$_2$(\%) \\
    \midrule
    $\tau$ (RQ1) & \textbf{80.9} & \textbf{81.4} & 81.6 & 81.6 & \textbf{83.5} & 80.5 & 89.8 & 87.8 \\
    \midrule
    &\multicolumn{8}{c}{Increasing minority classes} \\
    \cmidrule(lr){2-9}
    $\tau$+\textit{RO} & 79.5 & 75.1 & 90.1 & \colorbox{yellow!50}{\textbf{86.7}} & 79.1 & 74.2 & 91.2 & 87.2 \\
    $\tau$+\textit{BT} & 79.8 & 81.0 & 79.6 & 79.9 & 81.4 & 79.7 & 85.7 & 84.5 \\
    $\tau$+\textit{ER} & 79.6 & 77.1 & 86.1 & 84.1 & 79.8 & 79.2 & 82.7 & 81.9 \\
    $\tau$+\textit{NI} & 78.6 & 74.7 & 88.4 & 85.3 & 80.9 & 80.9 & 82.3 & 82.0 \\
    $\tau$+\textit{SR} & 80.4 & 79.7 & 83.0 & 82.3 & 79.1 & 80.7 & 78.2 & 78.7 \\
    $\tau$+\textit{All} & 79.1 & 76.8 & 85.4 & 83.5 & 79.6 & 76.3 & 87.8 & 85.2 \\
    \midrule
     &\multicolumn{8}{c}{Decreasing majority classes} \\
    \cmidrule(lr){2-9}
    $\tau$+\textit{RU} & 54.7 & 53.3 & \textbf{99.7} & 84.9 & 76.7 & 69.8 & \textbf{96.6} & \colorbox{yellow!50}{\textbf{89.7}} \\
    \midrule
     &\multicolumn{8}{c}{Increasing minority classes and decreasing majority classes} \\
    \cmidrule(lr){2-9}
    $\tau$+\textit{RO}+\textit{RU} & 63.0 & 58.4 & 98.3 & 86.5 & 82.5 & \textbf{85.0} & 81.7 & 82.3\\
    $\tau$+\textit{BT}+\textit{RU} & 58.1 & 55.2 & 99.3 & 85.6 & 77.5 & 71.3 & 94.6 & 88.8\\
    $\tau$+\textit{ER}+\textit{RU} & 60.7 & 56.9 & 98.3 & 85.8 & 79.3 & 79.7 & 80.3 & 80.2 \\
    $\tau$+\textit{NI}+\textit{RU} & 60.7 & 56.9 & 98.3 & 85.8 & 78.4 & 75.7 & 85.7 & 83.5 \\
    $\tau$+\textit{SR}+\textit{RU} & 59.3 & 56.0 & 99.0 & 85.8 & 80.4 & 75.4 & 91.8 & 88.0 \\
    $\tau$+\textit{All}+\textit{RU} & 61.6 & 57.4 & 99.0 & 86.5 & 78.4 & 74.9 & 87.4 & 84.6 \\
    \bottomrule
\end{tabularx}
  \begin{tablenotes}
     \item[$\dag$] $\tau$: Original training dataset, \textit{RO}: Random oversampling, \textit{RU}: Random undersampling, \textit{BT}: Back translation, \textit{NI}: Noise injection, \textit{ER}: Embeddings replacement, \textit{SR}: Synonym replacement, \textit{All}: All data augmentation methods, namely \textit{BT}, \textit{NI}, \textit{ER}, \textit{SR}. 
         \end{tablenotes}
\end{threeparttable}
\end{table*}

For binary classification, increasing the minority classes through random oversampling significantly improves the recall with a gain of 8.5 \textit{pp}, consequently F$_2$ with an gain of 5.1 \textit{pp}. All data augmentation methods also yield a better recall and F$_2$ than those achieved on the original dataset. The exception to this is \textit{BT} which applies  back translation techniques to triple the minority examples (in this case, the sentences that satisfy any provision). We believe that the low performance of \textit{BT} is because it (unlike other data augmentation methods) adds completely new sentences which might have lost the legal domain specificity due to translation. For instance, sentences that are written using ``shall'' modal verb to express a processor's obligation might be translated back to sentences without any modal verb, in which case the legal context concerning an obligation is no longer present. 
While increasing the minority classes did not improve accuracy and precision, the achieved values compared to the ones on the original dataset come with a loss ranging from -1.4 \textit{pp} to -0.5 \textit{pp} in accuracy and -6.3 \textit{pp} to -0.4 \textit{pp} in precision. 

Decreasing the majority classes alone or in combination with increasing the minority classes  yields a nearly perfect recall for binarcy classification with a notable improvement of $\approx$18 \textit{pp} over fine-tuning on the original dataset. However, this comes at the cost of huge decrease in precision of about $\approx$28 \textit{pp}. One reason is that minority classes for some provisions are very few and thus reducing the majority class would result in an overall smaller training dataset that is not informative enough to the binary classifiers. 
Combining the undersampling with increasing the minority classes does not help much in improving the precision. We conclude that the negative examples are particularly important to teach the classifier more distinguishable patterns to correctly predict the provisions in the DPA. This can be justified through our observation that the DPA text that satisfies any provision is highly similar to the text that does not satisfy any provision. Exposing both texts to the classifier is therefore helpful.

Unlike binary classification, decreasing the majority class for multi-class classification significantly improves recall and F$_2$ with a gain of $\approx$7 \textit{pp} and $\approx$2 \textit{pp}, respectively. Compared to binary classifiers which are fine-tuned for each provision, the multi-class classifier is exposed to all provisions as well as the other sentences that do not satisfy any provision. For this reason, reducing the majority class (\textit{other}, in this case) would have a positive effect. Nonetheless, precision has a loss of $\approx$12 \textit{pp}. We believe that this loss is still acceptable compared to the benefit of achieving a higher recall. 
Increasing the minority classes also improves the recall and F$_2$. In contrary to their behavior for binary classification,  data augmentation methods that work at the word level (e.g., SR and ER) performed much worse than \textit{BT} for the multi-class classification.  

We note that in our context data augmentation did not perform exceptionally well compared with random sampling techniques which can be implemented more efficiently. Therefore, we conclude that augmenting textual data is not very well suited for the legal domain due to the complexity of the legal language and the sensitivity of the concepts, terms, and words co-occurrences that can be disrupted through data augmentation methods.

\begin{tcolorbox}[arc=0mm,width=\columnwidth,
                  top=0mm,left=0mm,  right=0mm, bottom=0mm,
                  boxrule=1pt,colback=yellow!5!white,colframe=black]
In view of the above analysis, \textbf{the answer to RQ2 is} that handling the imbalance of the training dataset using \textit{RO} yields the best performing binary classifiers with an average F$_2$ of 86.7\% and applying \textit{RU} yields the best performing multi-class classifier with an F$_2$ of 89.7\%. 
\end{tcolorbox}

\sectopic{RQ3: How accurate are FSL solutions in checking the completeness of DPAs against GDPR?}
Table~\ref{tab:rq3} shows the results of three FSL scenarios for DPA completeness checking.  For comparison, we provide in the table the results of  the best performing solutions from RQ1 and RQ2, which are BERT for binary classification and RoBERTa for multi-class classification.

\begin{table*}[!t]
\caption{Accuracy Results for FSL-based Solutions \textbf{(RQ3)}. } \label{tab:rq3}
  \footnotesize

    \begin{threeparttable}[t]
    \begin{tabularx}{\textwidth}{@{}l*{10}{>{\centering\arraybackslash}X}@{}}
    \toprule
    & \multicolumn{4}{c}{Binary Classification } & \multicolumn{4}{c}{Multi-class Classification } \\
    \cmidrule(lr){2-5}\cmidrule(lr){6-9}
    & A(\%) & P(\%) & R(\%) & F$_2$(\%) & A(\%) & P(\%) & R(\%) & F$_2$(\%) \\
    \midrule
    RQ1 & \textbf{80.9} & 81.4 & 81.6 & 81.6 & \textbf{83.5} & 80.5 & 89.8 & 87.8 \\
    RQ2 & 79.5 & 75.1 & \textbf{90.1} & \colorbox{yellow!50}{\textbf{86.7}} & 76.7 & 69.8 & \textbf{96.6} & \colorbox{yellow!50}{\textbf{89.7}} \\
    \midrule
     &\multicolumn{8}{c}{FSL scenario 1: Fine-tuning LLMs on $30\%\subset\tau$, testing on $\epsilon$} \\
    \cmidrule(lr){2-9} 
     \ding{172} ALBERT & 54.4 & 68.9 & 21.1 & 24.5 & 75.3 & 82.3 & 66.3 & 69.0 \\
    \ding{173} BERT & 78.8 & 78.2 & 81.6 & 80.9 & 77.5 & 77.9 & 78.9 & 78.7 \\
    \ding{174} Legal-BERT & 67.5 & 82.6 & 46.9 & 51.4 & 78.6 & 82.3 & 74.5 & 75.9  \\
    \ding{175} RoBERTa & 67.0 & 79.8 & 48.3 & 52.4 & 79.8 & 82.5 & 77.2 & 78.2\\
        \midrule
    &\multicolumn{8}{c}{FSL scenario 2: Using FSL framework on $30\%\subset\tau$, testing on $\epsilon$} \\
    \cmidrule(lr){2-9}    
    \ding{181} SetFit & 77.7 & 77.6 & 79.9 & 79.4 & 79.6 & 80.5 & 79.9 & 80.0 \\
    \midrule
    &\multicolumn{8}{c}{FSL scenario 3: training ML on $30\%\subset\tau$, testing on $\epsilon$} \\
    \cmidrule(lr){2-9}
    \ding{176} BiLSTM & 78.8 & 80.1 & 78.2 & 78.6 & 78.2 & 78.7 & 68.0 & 69.9\\
    \ding{177} LR & 51.8 & 0.0  & 0.0  & 0.0 & 48.4 & \textbf{95.2} & 6.8  & 8.4 \\
    \ding{178} MLP &  69.1 & 86.6 & 52.9 & 57.4 & 71.5 & 89.3 & 45.6 & 50.5 \\
    \ding{179} RF & 58.2 & 82.4 & 28.6 & 32.9 & 60.0 & 80.4 & 25.2 & 29.2 \\
    \ding{180} SVM & 61.9 & \textbf{91.0} & 27.6 & 32.0 & 61.2 & 89.7 & 29.6 & 34.2\\
    \bottomrule
\end{tabularx}
\end{threeparttable}
\end{table*}

In the first scenario, we simply fine-tune the LLMs on 30\% of our dataset, corresponding to a total of $\approx$7200, of which 861 sentences satisfy any provision in GDPR. For binary classification, the table shows that LLM-based solutions perform much worse on the small proportion of the dataset except BERT. Comparing the performance on the original dataset (RQ1), BERT loses 0.7 \textit{pp} in F$_2$ when fine-tuned on  30\% of the data. This robust performance indicates that BERT is capable of learning effectively from few examples. 
The difference in performance of LLM-based solutions is more notable in the case of multi-class classification. For instance, RoBERTa has an loss of $\approx$10.5 \textit{pp} in F$_2$ compared to being fine-tuned on the dataset with its original size. Other LLMs do not perform well compared with their accuracy results reported in Table~\ref{tab:rq1.1}.

In the second scenario, we apply SetFit framework designed specifically for FSL. We use 30\% of the training dataset to have a better comparison with the other FSL scenarios. As explained in Section~\ref{sec:background}, SetFit leverages LLMs (concretely, sentence transformer) for  text classification. The framework attempts first to understand   the representations of the training examples (typically, sentences). This is done by fine-tuning the LLM on a few examples from the training data. The fine-tuned model is then used to generate the sentences embeddings that are the main elements for training a classifier tailored to the specific task. Compared with the best preforming solutions (BERT and RoBERTa in RQ2), SetFit provides reasonable results.
Specifically, SetFit has an average loss of $\approx$8.5 \textit{pp} in F$_2$.
However, for binary classification, fine-tuning BERT  on a smaller proportion of the dataset yields better results in our context than SetFit, with a gain of 1.5 \textit{pp} in F$_2$. On the contrary, SetFit performs slightly better in multi-class classification. 
The main advantage of FSL is that it requires much less labeled examples. With that in mind, the decision about whether to select SetFit or perform FSL depends highly on the application context. There is a clear trade-off between the manual effort needed to create labeled examples versus the accuracy of the solutions. 

In the last scenario, we train ML classifiers on 30\% of the dataset. In contrast with LLMs, the disadvantage of a smaller training dataset can be clearly seen in the table for ML-based solutions, reaching zero F$_2$  in the case of LR.      
The improvement of the performance of ML-based solutions when being trained on the original dataset confirms the necessity for large manually-labeled datasets for developing effective ML-based solutions. Even our training dataset which contains  about 3,400 positive examples was not enough for training accurate ML classifiers. 

\begin{tcolorbox}[arc=0mm,width=\columnwidth,
                  top=0mm,left=0mm,  right=0mm, bottom=0mm,
                  boxrule=1pt,colback=yellow!5!white,colframe=black]
\textbf{The answer to RQ3 is} that some LLMs such as BERT demonstrate a robust performance when fine-tuned on a small proportion of the training dataset (30\%, in our work). Using an FSL dedicated framework like SetFit yields reasonable accuracy, yet not better than fine-tuning LLMs on the entire dataset. The trade-off between manually-labeling more training examples versus the accuracy that is desired depends on the application context.   
\end{tcolorbox}
\sectopic{RQ4: What is the execution time of our proposed solutions?} 
We analyze the running time  considering both perspectives highlighted in Fig.~\ref{fig:approach}, namely user's and developer's perspectives.
To simply use our proposed alternative solutions, one can apply the readily-developed classifiers and run all the steps depicted in Fig.~\ref{fig:approach} except Step~3. Following the discussion in the previous RQs, we report below the time required for running four alternative solutions, namely BERT, RoBERTa, SetFit, and BiLSTM. The first two are the best performing solutions, whereas the latter two demonstrated a comparable performance in terms of accuracy as well as other advantages: SetFit is a perfect alternative when labeled data is scarce while BiLSTM is computationally efficient to develop compared to LLM-based solutions.

Considering the largest DPA in our our evaluation set with about 580 sentences, the first two steps to preprocess text and transform it into an intermediate representation required negligible time for all solutions.  
Steps~4~and~5 classify the text and check the completeness of the DPA. To do so, BERT and RoBERTa required $\approx5.5$ minutes and $\approx19$ seconds, respectively. Note that in the case of BERT, we run 19 different classifiers compared to one multi-class classifier in the case of RoBERTa. 
SetFit required $\approx6$ minutes and $\approx19$ seconds for respectively solving the binary and multi-class classification problems. Finally, BiLSTM required $\approx34$ seconds for solving the binary classification and $\approx1$ second for solving the multi-class classification. 

The developer must account for the time needed by step~3 to further develop the classifiers. Fine-tuning 19 BERT-based classifiers took about four weeks, whereas fine-tuning RoBERTa required about three days. Note that the fine-tuning time includes the hyper-parameter optimization. 
SetFit, which is designed to run on a CPU (instead of a GPU), took about 16 hours for developing the 19 binary classifiers and $\approx51$ minutes  for the multi-class classifiers. In comparison, BiLSTM took about two weeks to train and optimize the the hyper-parameter for binary classification and nearly one day for multi-class classification. 

\begin{tcolorbox}[arc=0mm,width=\columnwidth,
                  top=0mm,left=0mm,  right=0mm, bottom=0mm,
                  boxrule=1pt,colback=yellow!5!white,colframe=black]
\textbf{The answer to RQ4 is } that the estimated time for analyzing the completeness of a given DPA with 100 sentences ranges between 3 seconds and 1 minute when applying RoBERTa and BERT, respectively. This time is, on the one hand, practical from a user's perspective. Developing similar  Solutions, on the other, largely  depends on the availability of appropriate hardware (i.e., GPU) and sufficient number of labeled examples. Alternatively, one can develop SetFit, an FSL-based solution, or rely on BiLSTM, compromising thereby the overall accuracy.   
\end{tcolorbox}

\section{Threats to Validity} 
\label{sec:threats}
This section discusses the validity concerns pertinent to our work.

\sectopic{Internal Validity.} The main threat to internal validity concerns bias. To mitigate this threat, we constructed our labeled dataset with the help of three third-party annotators (non-authors). The annotators were not provided any details regarding the development of the different alternative solutions.

\sectopic{External Validity. } The main concern in respect of external validity is the generalizability of our proposed solutions. To this end, we note that we conducted our empirical evaluation on a large dataset of real DPAs covering different sectors. We believe that the obtained results reflect real-world conditions since the evaluation set was not exposed during the training or fine-tuning phases. More experimentation and user studies can be however beneficial for improving this validity concern.

\section{Conclusion} \label{sec:conclusion}
In this paper, we developed and evaluated ten alternative solutions for checking the completeness of data processing agreements (DPAs) against General Data Protection Regulation (GDPR). The alternative solutions utilize various enabling technologies, including traditional machine learning (ML), deep learning (DL), large-scale language models (LLMs) and few-shot learning (FSL). 
To evaluate our proposed solutions, we used a dataset of 163 real DPAs curated by three third-party (non-author) annotators. Our results indicate that the LLM-based solutions performed significantly better than ML-based solutions. The best performing solution is RoBERTa with an F$_2$ of 89.7\% for solving the completeness checking problem as a multi-class classification task. BERT performed equally well, with an F$_2$ score of 86.7\%, while enabling multi-label classification. Though these two solutions demonstrated effectiveness, our analysis highlights the huge benefits of two alternative solutions, including  BiLSTM (a DL algorithm) and  SetFit (an FSL framework). 
BiLSTM performs on par with LLM-based solutions but requires much less time to develop. SetFit yields less accurate results than LLMs, but it can be developed with much less training data. 
Overall, the results show that fine-tuning LLMs should be the most immediate option for addressing problems in the legal domain. However, selecting which LLM to choose plays a major role. For instance, though legal BERT was pre-trained on legal text, it was out-performed by other LLMS like RoBERTa which was exposed to general yet much larger text body. 
In the future, we would like to conduct user studies to assess the practical usefulness of these solutions that achieved promising results. 

\sectopic{Acknowledgment}
This paper was supported by Linklaters and Luxembourg's National Research Fund under grant BRIDGES/19/IS/13759068/ARTAGO. 

\sectopic{Declarations} 

\noindent\textit{Data Availability: }
We make all our non-proprietary material used in our empirical evaluation publicly available at this link

\href{https://figshare.com/s/77338e558ffb6adf6f55}{https://figshare.com/s/77338e558ffb6adf6f55}. 

\noindent\textit{Conflict of Interest: }
The authors declared that they have no conflict of interest. 


\bibliographystyle{spmpsci}
\bibliography{paper}
\end{document}